\documentstyle[12pt]{article}
\newlength{\dinwidth}
\newlength{\dinmargin}
\newcommand{\ba}{\begin{array}}
\newcommand{\ea}{\end{array}}
\newcommand{\be}{\begin{equation}}
\newcommand{\ee}{\end{equation}}
\newcommand{\bea}{\begin{eqnarray}}
\newcommand{\eea}{\end{eqnarray}}

\def\bee{\begin{eqnarray}}
\def\eee{\end{eqnarray}}
\def\be{\begin{equation}}
\def\ee{\end{equation}}

\begin{document}
\thispagestyle{empty}
\addtocounter{page}{-1}
\begin{flushright}
IASSNS-HEP 97/36\\
SNUTP 97-051\\
{\tt hep-th/9704158}\\
\end{flushright}
\vspace*{1.3cm}
\centerline{\Large \bf Heterotic M(atrix) Strings and Their Interactions
\footnote{
Work supported in part by the NSF-KOSEF Bilateral Grant,
KOSEF Purpose-Oriented Research Grant 94-1400-04-01-3 and SRC-Program, 
Ministry of Education Grant BSRI 97-2410, the Monell Foundation
and the Seoam Foundation Fellowships.}}
\vspace*{1.2cm} \centerline{\large\bf Soo-Jong Rey }
\vspace*{0.8cm}
\centerline{\large\it School of Natural Sciences, Institute for
Advanced Study}
\vskip0.15cm
\centerline{\large\it Olden Lane, Princeton NJ 08540 USA}
\vskip0.3cm
\centerline{\large \it Physics Department, Seoul National University, 
Seoul 151-742 KOREA}
\vspace*{2.1cm}

\centerline{\Large\bf Abstract}
\vskip0.5cm
Following recent proposal of Dijkgraaf, Verlinde and Verlinde, we show that
the M(atrix) theory compactified on ${\bf S}_1/{\bf Z}_2$ provides with a
non-perturbative description of second-quantized light-cone heterotic string.
This so-called heterotic M(atrix) string theory is defined by two-dimensional
(8,0) supersymmetric chiral gauge theory with gauge group SO(2N) in the large
N limit. We argue that at strong coupling fixed point the chiral gauge theory
flows to a (8,0) superconformal field theory defined via $S_N$ symmetric
product space orbifold. We show that the leading order correction to the
strong coupling expansion corresponds to a unique irrelevant operator of
scaling dimension three and describes joining and splitting cubic interactions
of light-cone heterotic string. We also speculate on M(atrix) description of
bosonic strings via dimensional reduction of d=26 Yang-Mills theory.
\vspace*{1.1cm}

\centerline{Submitted to Nuclear Physics B}

\newpage

%%%%%%%%%%%%%%%%%%%%%%%%%%%%%%%%%%%%%%%%%%%%%%%%%%%%%%%%%%%%%%%%%%%%%%%%
\section{Introduction}
%%%%%%%%%%%%%%%%%%%%%%%%%%%%%%%%%%%%%%%%%%%%%%%%%%%%%%%%%%%%%%%%%%%%%%%%
The eleven-dimensional M theory~\cite{witten} 
is a unifying theory of all known
superstring theories in the strong coupling limit.
Apart from the fact that massless excitations
are described by the eleven-dimensional supergravity, however, little has been
understood as to what the underlying fundamental constituents of the
M theory are and what governs microscopic dynamics of the constituents.
In an attempt to gain better understanding, one may draw a hint from the 
history of probing internal structure of hadrons. Partons inside strongly 
coupled hadrons are very fuzzy due to
their characteristic high-frequency motion.  If the hadrons are boosted,
the motion is slowed down via Lorentz time dilation, hence, makes it
possible to snapshot the parton strucutre. Indeed, morally following this
idea, Banks {\sl et.al.}~\cite{bfss}
 have proposed a partonic definition of the M theory. In the infinite
momentum limit, light-front view of a strongly coupled type IIA string is
infinitely many zero-branes threaded on the string itself.
One thus discovers that M-theory partons 
consist of zero-branes and infinitely short open strings gluing them
together. As such the M theory parton dynamics is accurately described by
large N limit of ${\cal N} = 16$ supersymmetric $U(N)$ matrix quantum
mechanics, the so-called M(atrix) theory.
Despite clear identification of fundamental constitutent partons and
underlying dynamics governing them, it has been very difficult to
extract intrinsically M theoretic physics, for example, nontrivial
S-matrix amplitudes among physical asymptotic states.

Very recently Dijkgraaf, Verlinde and Verlinde (DVV)~\cite{dvv}
have offered an important new insight to the M(atrix) theory. 
By compactifying the M(atrix) theory one more dimension on ${\bf S}_1$ in 
addition to the ${\bf S}_1$ compactified `quantum' dimension and utilizing
exchange symmetry between the two directions, 
DVV have convincingly
argued that the resulting M(atrix) string theory provides for a 
nonperturbative description of M theory compactified on ${\bf S}_1$. 
In particular, in the weak coupling, 
DVV have shown that, based on earlier basic observation by 
Motl~\cite{motl} and identification of moduli space by 
Banks and Seiberg~\cite{banksseiberg}, 
the M(atrix) string theory describes second-quantized 
light-cone type IIA string, Virasoro projection for individual strings
emerges
from residual ${\bf Z}_N$ discrete gauge symmetry in the large N limit
and, most importantly, 
their joining and splitting interaction vertices. The DVV observation 
has marked a significant progress since it has promoted the original M(atrix) 
theory into a firmer and calculable set-up.

On the other hand, it is not obvious that DVV proposal is robust 
enough and extendible to all other superstrings than type IIA.
In M theory, different string theories arise from different 
choices of the compactification manifold.
For instance, M theory compactified on an orbifold ${\bf S}_1/{\bf Z}_2$ 
yields heterotic and type I${}^\prime$ superstrings. As they have different 
symmetries and field content on the worldsheet, in particular, half many
spacetime supersymmetries compared to type II strings,
it provides a highly nontrivial check point to test whether the DVV
proposal to the M(atrix) theory applies successfully to other superstrings
as well.

In this paper, with the above motivation, we study the proposal of DVV
for heterotic superstring.  We show that heterotic M(atrix) 
theory~\cite{swedish, kimrey} 
compactified on a circle defines the heterotic
M(atrix) string theory in terms of $(8,0)$ supersymmetric chiral
gauge theory coupled to twisted sector matter multiplets. 
In the strong coupling limit we find that the theory reproduces 
the spectra and interactions of light-cone Green-Schwarz 
heterotic superstring. We also explore the possibility of M(atrix)
theory for bosonic strings in terms of dimensionally reduced d=26
Yang-Mills gauge theory.
While this paper was being prepared, we have received preprints
related to part of our results~\cite{banksmotl}~\cite{lowe}.

%%%%%%%%%%%%%%%%%%%%%%%%%%%%%%%%%%%%%%%%%%%%%%%%%%%%%%%%%%%%%%%%%%%%%%%%%%%%
\section{Heterotic M(atrix) String Theory}
%%%%%%%%%%%%%%%%%%%%%%%%%%%%%%%%%%%%%%%%%%%%%%%%%%%%%%%%%%%%%%%%%%%%%%%%%%%%
In the previous work~\cite{kimrey}, we have obtained 
the heterotic M(atrix) theory by compactifying the 
M(atrix) theory on an orbifold ${\bf S}_1/{\bf Z}_2$ along the 11-th
direction. Following the proposal of DVV, we compactify the heterotic 
M(atrix) theory further on ${\bf S}_1$ along, say, the 9-th direction.
From the defining M(atrix) theory~\cite{bfss} point of view, 
we have compactified the theory
on a cylinder ${\bf S}_1 \times {\bf S}_1/{\bf Z}_2$.  
We now interchange 9-th and 11-th directions. This results
in the heterotic M(atrix) string theory, a M(atrix) theory description 
of second-quantized heterotic strings. Because of $11-9$ direction flip, 
the heterotic string is boosted longitudinally and $p_+$ momentum is 
measured in units of $1/R_{11}$, the same unit that was used before 
for D0-brane particle number. Likewise, string coupling is now determined 
by the 9-th direction radius $R_9$. 
As such, in the M(atrix) heterotic string theory, the 
heterotic strings are described naturally 
in the light-cone Green-Schwarz formulation.

The correspondence between the heterotic string and arrays of D0-branes
through $11-9$ direction flip can be understood via a chain of 
by-now well-established $S-$ and $T-$ dualities. Consider
the D-particles arrayed on 11-th direction. T-duality along 9-th direction
turns the D-particles into D-strings in type I string theory. S-duality
(residual ${\bf Z}_2$ duality of the underlying type IIB 
$SL(2, {\bf Z})$ duality)
converts the D-string into heterotic string. Inverting 
the T-duality along 11-th direction, we have arrived at heterotic 
strings with longitudinal momentum around 11-th direction. The mapping
is depicted in the following diagram:
\vskip0.5cm
\be
\begin{array}{ccccc} {\bf Type~ I^\prime~ D0-brane~ no. = N} & 
\leftarrow & ({\tt T-duality}) & \rightarrow &
{\bf Type~ I~ D1-string~ no. = N} \\
\uparrow & & & & \uparrow \\
({\tt 11-9~  flip}) & & & &
({\tt S-duality}) \\
\downarrow & & & & \downarrow \\
{\bf Heterotic~ momentum~ P_+ = N} &
\leftarrow & ({\tt T-duality}) & \rightarrow &
{\bf Heterotic~ winding~ no. = N} \end{array}
\ee
\vskip0.5cm

In this section, we elaborate details of the heterotic M(atrix) string
theory construction along the lines sketched above. As we will see,
the resulting theory is an $(1+1)$-dimensional $(8,0)$ supersymmetric
gauge theory defined on $ {\bf S}_1 \times {\tilde {\bf S}}_1 $, 
in which the orbifold has turned into a circle ${\tilde {\bf S}}_1$ of 
radius $1/R_9$.

%%%%%%%%%%%%%%%%%%
\subsection{Heterotic M(atrix) Theory}
%%%%%%%%%%%%%%%%%%%%%%
We begin by reviewing aspects of the heterotic M(atrix) theory
relevant for foregoing discussions. By definition, heterotic M(atrix)
theory is the M(atrix) theory compactified on an orbifold $I =
{\bf S}_1/ {\bf Z}_2$, say, in 9-th direction~\cite{swedish, kimrey}. 
In the previous paper~\cite{kimrey}, we have 
studied constraint of the orbifold condition and have shown that
N D0-brane parton dynamics is governed by ${\cal N}= 8$ supersymmetric
SO(2N) matrix quantum mechanics. The orbifold compactification breaks
the R-symmetry to Spin(8) $\subset$ Spin(9).
As such, we adopt a Majorana spinor convention so that real and symmetric
representations of the Spin(9)
gamma matrices $\Gamma_I, (I = 1, \cdots, 9)$ are decomposed into
\be
\Gamma_i = \left( \begin{array}{cc} 0 & \sigma^i_{a \dot a} \\
\sigma^i_{{\dot a} a} & 0 \end{array} \right)
\hskip0.5cm i = 1, \cdots, 8; 
\hskip0.75cm \Gamma_9 = \left( \begin{array}{cc} 
-\delta_{ab} & 0 \\ 0 & +\delta_{{\dot a} {\dot b}} \end{array} \right)
\ee
in terms of Spin(8) gamma matrices $\sigma^i$'s.
The Spin(9) spinor\footnote{We adopt the conjugate spinor convention
${\overline {\bf \Theta}} \Gamma_- = {\bf \Theta}^T$.}
$\bf \Theta$ is decomposed into two inequivalent chiral spinors of Spin(8):
\be
{\bf \Theta} = {\bf 8}_{\rm s} \oplus {\bf 8}_{\rm c} \equiv {\bf S}_a \oplus 
{\bf S}_{\dot a}
\ee
where 
\be
\Gamma_9 {\bf S}_a = - {\bf S}_a ; \hskip1cm 
\Gamma_9 {\bf S}_{\dot a} = + {\bf S}_{\dot a}.
\ee
The field content of the heterotic M(atrix) theory consists of untwisted
and twisted sectors and has been determined as follows~\cite{kimrey}.
Untwisted sector consists of a non-dynamical gauge supermultiplet $(A_0; 0)$
with purely bosonic degrees of freedom, an adjoint supermultiplet
$(A_9, {\bf S}_{a})$ and a rank-2 symmetric supermultiplet 
$({\bf X}_i, {\bf S}_{\dot a})$.
 
Twisted sector states arise from ${\bf S}_1/{\bf Z}_2$ orbifold fixed points.
Located at each of the two fixed points $X_9 = 0, \pi R_9$ are sixteen units of
Ramond-Ramond nine-form charges.  
To cancel these anomalous charges, we introduce fundamental
supermultiplets  $(0; \chi_A^{(1)})$ and $(0; \chi_B^{(2)})$ 
$(A,B = 1, \cdots, 16)$\footnote{Note that purely bosonic adjoint and
purely fermionic fundamental supermultiplets are compatible with 
${\cal N} = 8$ supersymmetric matrix quantum mechanics with
$Spin (8)$ R-symmetry.} representing two sets of sixteen D8-branes.
In order to cancel the Ramond-Ramond gauge flux {\sl locally}, 
it is necessary to lock the D8-brane positions on top of the
orbifold fixed points. This is achieved by turning on Wilson line of the
spacetime gauge field $B^9_{AB} = \pi R_9 \, {\rm diag}
(0, \cdots, 0 , 1, \cdots, 1) \otimes i \sigma_2$, where $i \sigma_2$
acts on each pairs of D8-brane and its mirror image.
This Wilson line configuration breaks the heterotic gauge group 
$E_8 \times E_8$ or Spin$(32)/{\bf Z}_2$ to $SO(16) \times SO(16)$.
In what follows, we thus restrict our discussions to
$G \equiv SO(16) \times SO(16)$ gauge group configuration.

The complete spectra and their quantum numbers of the heterotic M(atrix)
theory are summarized in the following table.
\vskip0.5cm
\begin{tabular}{|c|c|c|c|c|c|} \hline
Sector & Multiplet & Bosons & Fermions & Spin(8) & SO(2N) \\ \hline
untwisted & gauge & $A_0$ & $\cdot$ & $({\bf 1} \,\,;\,\, 0)$
 & 2N(2N--1)/2 \\ \hline
 & adjoint & $A_9$ & ${\bf S}_{ a}$ & $({\bf 1}\,\,; \,\,{\bf 8}_{\rm s})$
& 2N (2N--1)/2
 \\ \hline
 & symmetric & ${\bf X}^i$ & ${\bf S}_{\dot a}$ 
& $({\bf 8}_{\rm v} \,\,;\,\, {\bf 8}_{\rm c})$ &
2N(2N+1)/2
\\ \hline
twisted & fundamental & $\cdot$ & $\chi^{(1,2)}_M$ & $(\,\,0\,\,;\,\, {\bf 1})$
& 2N
\\ \hline
\end{tabular}
\vskip0.5cm

Recalling that the two sets of sixteen twisted sector fermions are locked 
at $X_9 = 
0, \pi R_9$ symmetrically by turning on the Wilson line $B^9$, we find that
the heterotic M(atrix) theory is defined by the Lagrangian:
\bee
L = {\rm Tr} \, \Big(\!\!\!\!\! &-& \!\!\!\!\!
{1 \over 2 R_{11}} (D_\tau A_9)^2 
+ {1 \over 2 R_{11}} (D_\tau {\bf X})^2 
- {R_{11} \over 2} [A_9, {\bf X}_i]^2
+ {R_{11} \over 4} [{\bf X}^i, {\bf X}^j]^2
\nonumber \\
&-& {\bf S}_{ a} D_\tau {\bf S}_{ a} + R_{11} \, {\bf S}_{ a}
[A_9, {\bf S}_{a}]
+ {\bf S}_{\dot a} D_\tau {\bf S}_{\dot a} + R_{11} \, {\bf S}_{\dot a}
 [A_9, {\bf S}_{\dot a}]
- 2R_{11} \, {\bf X}^i \sigma^i_{a \dot a}
\{ {\bf S}_a, {\bf S}_{\dot a} \} \Big)
\nonumber \\
&+& \chi^{(1)}_A (D_\tau + R_{11} \, A_9) \chi^{(1)}_A
+  \chi^{(2)}_A (D_\tau + R_{11} \, (A_9 - i \pi R_9 \sigma_2 )) \chi^{(2)}_A.
\label{hetlag}
\eee
Here, $D_\tau \equiv 
\partial_\tau - [A_0,\,\,\, ]$ defines the covariant derivative.
The normalization of the twisted sector fermions to the $A_9$ 
gauge multiplet has been fixed so that the mass scale of the
open string, represented by the twisted sector fermions,
 connecting the D0-brane parton and the D8-brane
has the same mass scale as the open string among the D0-branes.
The common mass scale is the prerequisite to the cancellation of
induced one-loop vacuum energy~\cite{kimrey} as well as quantum
mechanical ${\bf Z}_2$ global anomaly~\cite{banksseibergsilverstein}.

The twisted sector is well-defined only if the fermion numbers are even.
This is because the one-dimensional Dirac operator $i D_\tau
= i d / d \tau + i A_0$ is not an elliptic operator. From the D8-brane
point of view, this conditions is automatically guaranteed since there
exists always a mirror D8-brane for every D8-brane. 
From the covering space point of view,
this implies that the heterotic M(atrix) theory is defined, in addition
to SO(2N) gauge group, with
$[{\bf Z}_2]^N$ bundle which acts trivially on the untwisted sector 
but nontrivially on the twisted sector fermions. The action of 
$[{\bf Z}_2]^N$ bundle is then identified with 
\be
[{\bf Z}_2]^N: \hskip0.75cm 
\chi^{(1,2)}_{Ap}  \rightarrow {\eta_p}^q \, \chi^{(1,2)}_{Aq}, 
\hskip1cm 
{\eta_p}^q = {\rm diag.} (\pm, \pm, \cdots, \pm)_{ N \times N}.
\label{z2gaugesymmetry}
\ee 

From the above Lagrangian, we find the light-cone Hamiltonian as:
\bee
H_{\rm LC} = R_{11} \, \Big[ {\rm Tr} \, \Big(\!\! \!\!\! &-& \!\!\!\!\! 
{1 \over 2} \Pi_9^2  + {1 \over 2} {\bf \Pi}^2_i
+ {1 \over 2} [A_9, {\bf X}^i]^2 - {1 \over 4} [{\bf X}^i, {\bf X}^j]^2
\nonumber \\
&-& {\bf S}_{\dot a} [A_9, {\bf S}_{\dot a}] - {\bf S}_{a}
[A_9, {\bf S}_a] + 2 {\bf X}^i \sigma^u_{a \dot a}
\{ {\bf S}_a, {\bf S}_{\dot a} \} \Big)
\nonumber \\
&-& \chi^{(1)}_A A_9 \chi^{(1)}_A - \chi^{(2)}_A (A_9 - i \pi R_9 \sigma_2 ) 
\chi^{(2)}_A \, \Big].
\label{hetham}
\eee

The eight kinematical generators ${\bf Q}^{\dot a}$
 and the eight dynamical ones ${\bf Q}^a$ are given by:
\bee
{\bf Q}^{\dot a}  &=& { 1 \over {\sqrt {R_{11}} }}
{\rm Tr} {\bf S}^{\dot a} , \nonumber \\
{\bf Q}^a &=& {\sqrt {R_{11}} } {\rm Tr} 
\Big( (\sigma^i_{a \dot a} {\bf S}^{\dot a} {\bf \Pi}^i - {\bf S}^a 
\Pi_9 ) + {1 \over 2}
( \sigma^{ij}_{ab} {\bf X}^i [ {\bf S}^b, {\bf X}^j]
+ \sigma^i_{a \dot a} {\bf S}_{\dot a} 
[A_9, {\bf X}^i]) \Big).
\eee
It is straightforward to check that the anticommutators of the 
${\bf Q}^{\dot a}, {\bf Q}^a$
supercharges give rise to the conserved longitudinal momentum
$P_+ = N / R_{11}$ and the light-cone Hamiltonian $H_{\rm LC}$ 
up to Gauss' law constraint respectively.
%%%%%%%%%%%%%%%%%%%%%%%%%%%%%%%%%%%%%%%%%%%%%%%%%%%%%%%%%%%%%%
\subsection{Heterotic M(atrix) String Theory}
%%%%%%%%%%%%%%%%%%%%%%%%%%%%%%%%%%%%%%%%%%%%%%%%%%%%%%%%%%%%%%
We now construct the heterotic M(atrix) string theory adopting the DVV
proposal. From the M-theory point of view, the heterotic
and the Type I strings arise from two different degeneration limits of
a cylindrical membrane stretched between the two orbifold fixed points 
along 9-th direction. In the limit the 11-th circular direction shrinks,
the membrane is reduced to the Type I string with one-dimensional
Chan-Paton fermions at each ends. 
In the limit the 9-th orbifold direction shrinks, the membrane
becomes heterotic string with chiral fermions generating Kac-Moody
current algebra. In particular,
from Type I string
point of view, the heterotic string is nothing but the strong-weak
coupling dual D-string~\cite{polchinskiwitten}.  

The heterotic string is properly described by dualizing 
along the squeezed, 9-th orbifold direction~\cite{kimrey}. 
We first note that 
T-duality maps ${\bf S}_1/{\bf Z}_2$ orbifold with radius $R_9$ into 
${\tilde {\bf S}}_1$ circle with dual radius $1/R_9$. This is easy
to understand from the D-brane and orientifold configurations. Initial
configuration is given by 8-orientifolds and sixteen D8-branes located
at the two fixed points $X_9 = 0, \pi R_9$. T-duality along the
9-th direction turns them into 9-orientifold and D9-branes.
Since they have co-dimensions zero in the transverse space, the 
9-orientifolds and D9-branes do not give rise to any orbifolding at all. 
In particular, 9-th direction after T-duality becomes ${\tilde {\bf S}}_1$
of radius $1/R_9$.
We thus introduce a worldsheet parameter $\sigma \in [0, 2 \pi]$ so that
the distance along the dualized 9-th orbifold direction is measured
by $\sigma/R_9$. With this convention, the duality turns 
$A_9 \rightarrow R_9 D_\sigma$ and $\Pi_9 \rightarrow E_\sigma / R_9$
in Eqs.~(\ref{hetlag},~\ref{hetham}).
This mapping turns the untwisted sector of the ${\cal N}=8$ 
supersymmetric SO(2N) matrix quantum mechanics 
Eqs.~(\ref{hetlag},~\ref{hetham}) 
into $(8,0)$ supersymmetric SO(2N) chiral gauge theory.

For twisted sector fermions, the T-duality is achieved by replacing
$A_9 \rightarrow R_9 D_\sigma$. The transformation makes the twisted
sector fermions propagate chirally 
around ${\tilde {\bf S}}_1$. We thus treat them 
symmetrically by absorbing the $\pi R_9$ dependence of $\chi^{(2)}$ and 
introducing redefined chiral fermions $\chi^{(P,A)}$    :
\bee
\chi^{(P)}_A (\sigma + \tau) &\equiv& \chi^{(1)}_A (\sigma + \tau),
\nonumber \\
\chi^{(A)}_A (\sigma + \tau)  &\equiv& e^{ i \sigma/2} \, \chi^{(2)}_A
(\sigma + \tau).
\eee
Hence, $\chi^{(P,A)}$ have opposite boundary conditions each
other once
traversed around ${\tilde {\bf S}}_1$, $\sigma \rightarrow \sigma
+ 2 \pi$. This is the M(atrix) string theory manifestatation that 
distinguishes the two sets of twisted sector fermions coming from 
each orbifold fixed points at $A_9 = 0, \pi R_9$. We adopt a  
convention in which $\chi^{(P)}$ is periodic and $\chi^{(A)}$ is 
anti-periodic. In fact, the two sets of boundary conditions, hence,
twisted sector fermions mix each other. In addition, in 
the heterotic M(atrix) theory, we have identified the $[{\bf Z}_2]^N$ 
discrete gauge symmetry acting only on the twisted sector fermions, 
Eq.~(\ref{z2gaugesymmetry}). Upon T-duality to heterotic M(atrix) 
string theory, the $[{\bf Z}_2]^N$
discrete gauge symmetry then acts together with the above 
boundary conditions. Altogether, this defines the action of
nontrivial SO(2N)$\times [{\bf Z}_2]^N$ bundle via `t Hooft's
twisted boundary conditions on the twisted sector fermions.

We thus find the heterotic M(atrix) string theory consisting of the
following field content and quantum numbers:
\vskip0.5cm
\begin{tabular}{|c|c|c|c|c|c|} \hline
Sector & Multiplet & (Components) & Spin(8) & SO(2N) 
& Worldsheet  \\ \hline
untwisted & gauge & $(A_\tau, A_\sigma$ ; ${\bf S}_{\dot a}$) & 
$({\bf 1} \, ; \, {\bf 8}_{\rm c})$  & 2N (2N--1)/2 & left-moving
 \\ \hline
 & symmetric & $({\bf X}^i$; ${\bf S}_a)$ & $({\bf 8}_{\rm v}; 
{\bf 8}_{\rm s}) $ &  2N(2N+1)/2 & right-moving
\\ \hline
twisted & fundamental & (\,\,$\cdot$\,\, ; $\chi^{(P,A)}_A$)
 & $(0 \,\, ; \,\, {\bf 1})$ & 2N & left-moving
\\ \hline
\end{tabular}
\vskip0.5cm

To comply with
the conventional light-cone string worldsheet units, 
we make a few more rescalings together with the T-duality. First, 
we convert the M-theory time
into worldsheet time by rescaling $t \rightarrow \tau / R_{11}$. Next,
we normalize the kinetic terms canonically by rescaling ${\bf X}^i 
\rightarrow {\bf X}^i/{\sqrt {R_9}}$. Finally, via 11-9 flip, we
identify the heterotic string coupling parameter 
$g_H^2 \equiv R_9^3$ in M-theory unit.

After the rescaling, the action is given by:
\bee
S = \int \! d\tau \oint_0^{2 \pi} {d \sigma \over 2 \pi}
\Big[\, {\rm Tr} \Big(\!\!\! &+&\!\!\! {g_H^2 \over 4 } F^2_{\alpha \beta}
+ {1 \over 2} (D_\alpha {\bf X}^i)^2 + {1 \over 4 g_H^2}
[{\bf X}^i, {\bf X}^j]^2
\nonumber \\
&+& {\bf S}_a {\cal D}_R {\bf S}_a + 
{\bf S}_{\dot a} {\cal D }_L  {\bf S}_{\dot a} 
- 2 {\bf X}^i \sigma^i_{a \dot a}
\{ {\bf S}_a, {\bf S}_{\dot a} \} \Big)
\nonumber \\
&+&  \chi_A^{(P)} {\cal D}_L
\chi_A^{(P)}
+  \chi_B^{(A)} {\cal D}_L \chi_B^{(A)} \Big].
\label{hetmatlag}
\eee
Here, the left-, right-moving covariant derivatives are defined as
${\cal D}_{ L, R} \equiv D_\tau  \pm D_\sigma$.

The Hamiltonian of the heterotic M(atrix) string theory is given by
\bee
H_{\rm LC} =  \oint_0^{2 \pi} {d \sigma \over 2 \pi}
\Big[ \, {\rm Tr} 
\Big(\!\!\! &+&\!\!\! {1 \over 2 g_H^2} E^2 + {1 \over 2} {\bf \Pi}^2_i
+ {1 \over 2} (D_\sigma {\bf X}^i)^2 - {1 \over 4 g_H^2} 
[{\bf X}^i, {\bf X}^j]^2
\nonumber \\
&-& {\bf S}_{\dot a} D_\sigma {\bf S}_{\dot a}
- {\bf S}_a D_\sigma {\bf S}_a
+ {1 \over g_H }\, 2 {\bf X}^i \sigma^i_{a \dot a}
\{ {\bf S}_a, {\bf S}_{\dot a} \} \Big)
\nonumber \\
&-& \chi^{(P)}_A D_\sigma \chi^{(P)}_A
- \chi^{(A)}_B D_\sigma \chi^{(A)}_B \Big].
\label{hetmatham}
\eee
In the adopted normalization convention, the longitudinal momentum is
taken as $p_+ = 1$ for {\sl all} $N$. 
That this corresponds to the heterotic string can be seen~\cite{kimrey}
by formally
taking $N = 1/2$. Then, the Lagrangian is reduced precisely to the
same form as the worldsheet action of a single 
light-cone Green-Schwarz heterotic string, the structure first identified 
by Polchinski and Witten~\cite{polchinskiwitten} from the
D-string worldsheet action in Type I string theory via
 ${\bf Z}_2$ heterotic -- Type I duality.
Therefore, for $N > 1$, it is expected that the heterotic M(atrix) 
theory contain N independent heterotic strings once suitably interpreted.
Indeed, in the next Section, following the DVV proposal, we will
show that the
heterotic M(atrix) theory indeed provides for a new second-quantization
description of light-cone Green-Schwarz heterotic strings. 

The resulting heterotic M(atrix) string theory is a $(1+1)$-dimensional
chiral gauge theory. Spectrum of the theory is tightly constrained by
the requirement of anomaly cancellation. From the identified spectrum
above, 
we find that the gauge anomaly from the right-moving symmetric
multiplet fermions is cancelled by the left-moving gauge and fundamental
twisted sector fermions. In fact, the anomaly cancellation requirement
may be considered as the M(atrix) theory principle for identifying the
twisted sector spectrum~\cite{banksseibergsilverstein,kimrey}. The 
result is also consistent with the twisted sector spectrum derived from
one--loop vacuum energy cancellation requirement in the heterotic 
M(atrix) theory, as is expected from the T-duality between the two 
theories~\footnote{It should be noted that the d=10, ${\cal
N} = 1$ supersymmetric Yang-Mills theories other than SO(32)
gauge group are anomalous, hence, inconsistent. The 
M(atrix) theory governing D0-brane parton dynamics has been derived 
from this theory via dimensional reduction. Given that ${\bf T}^d$
toroidal compactification is described by $(d+1)$-dimensional 
supersymmetric Yang-Mills theory, the M(atrix) 
theory may encounter an inconsistency once all nine dimensions are 
compactified. Such potential problem does not arise if higher dimensional
toroidal compactifications of M(atrix) theory are instead
described by some
other field theories than supersymmetric Yang-Mills theories. 
In fact, there exist such indications from the $(2,0)$ strong coupling 
fixed points for ${\bf T}^4$ and ${\bf T}^5$ 
compactifications~\cite{berkoozrozali}. We expect that such a resolution
continues to higher-dimensional toroidal compactifications. Alternative
possibility is that the gauge group for ${\bf T}_9$ compactification 
is not enhanced to $U(N)$ but manifests $[U(1)]^N$ only.}   

A remark is in order. One may question possible higher--order corrections
to the above Lagrangian and Hamiltonian 
Eqs.~(\ref{hetmatlag},~\ref{hetmatham}),
especially, higher-dimensional operators involving gauge field strengths.
However, we believe that there are no such terms. The
Hamiltonian Eq.~(\ref{hetmatham}) has the full structure of
the light-cone Green-Schwarz heterotic string already for $N=1/2$ as 
shown above. The quadratic terms in Eq.~(\ref{hetmatham}) saturates
all we can have consistently with light-cone kinematics and supersymmetry.
As will be shown in the next section, this is also the
case for arbitrary $N$ in the phase the gauge group is spontaneously 
broken to $S_N \times [{\bf Z}_2]^N $. 
  
%%%%%%%%%%%%%%%%%%%%%%%%%%%%%%%%%%%%%%%%%%%%%%%%%%%%%%%%%%%%%%%%%%%%%%
\section{Phases of Heterotic M(atrix) String Theory}
%%%%%%%%%%%%%%%%%%%%%%%%%%%%%%%%%%%%%%%%%%%%%%%%%%%%%%%%%%%%%%%%%%%%%%
In the previous section, we have defined the heterotic M(atrix) string
theory in terms of $(1+1)$-dimensional
 $(8,0)$ supersymmetric chiral gauge theory.
In this section, following the idea of \cite{dvv}
we show that the heterotic M(atrix) string defines
the second-quantized heterotic strings by analyzing the phases
of the supersymmetric chiral gauge theory.

The basic results may be summarized as follows. 
We will first show that the strong coupling Higgs phase of $SO(2N)$
chiral gauge theory defines non-interacting, light-cone multi-heterotic 
string states with total number of string $\le N$ and the total 
longitudinal momentum $p_+ = N/R_{11}$. In addition, we show that
string joining and splitting interactions are provided precisely
by the strong coupling expansions of the chiral gauge theory. 

\subsection{Free Heterotic String Limit}
The gauge coupling constant $g_{\rm YM}$ in (1+1)-dimensions
has a mass dimension 1. 
Since $g_{\rm YM}^{-2}  = \ell_s^2 g_H^2$, the dimensionless 
coupling governing the gauge dynamics is given by 
${\tilde g}_{\rm YM}^2 = (\Sigma / \ell_s^2) g_H^{-2}$ where 
$\Sigma$ denotes the world-sheet area.   

We first consider the strong gauge coupling limit, ${\tilde g}_{\rm YM} 
\rightarrow \infty$. 
This is the limit for which $g_H \rightarrow 0$ and $\Sigma \rightarrow
\infty$. The strongly coupled heterotic M(atrix) string
theory flows to an infrared fixed point at which the theory is 
described by a nontrivial conformal field theory.
Content of the conformal field theory is tightly constrained by 
symmetries. The
(8,0) chiral supersymmetry, Spin(8) R-symmetry and $G \equiv 
SO(16) \times SO(16)$ affine Lie algebra on the 
worldsheet should be respected by a
candidate conformal field theory~\footnote{This can be argued from
`t Hooft anomaly matching conditions.}. 
We claim that such a conformal field theory is provided by 
the $(8,0)$ supersymmetric sigma model on the symmetric product 
space orbifold:
\be
S^{N} ({\bf R}^8 \otimes G )
=  ({\bf R}^8 \otimes G )^{N} /(S_{N} \times [{\bf Z}_2]^N)
\label{symmetricspace}
\ee
where $G = SO(16) \times SO(16)$. Configuration of the other $N$--image
strings is isomorphic
to Eq.~(\ref{symmetricspace}) so we do not repeat them in what follows.
It is straightforward to identify
the above target space. In the free heterotic string 
limit $g_H \rightarrow 0$, the ${\bf X}^i$ fields are dynamically 
confined on the moduli space ${\cal M}_N \equiv \{ {\bf X}^i: 
[{\bf X}^i, {\bf X}^j] = 0 \}$.
At the same time, the gauge supermultiplet $(A_0, A_9; {\bf S}_{a})$ 
decouples, leaving only 
charge neutrality constraint for the remaining
symmetric and fundamental multiplets. 
At generic point in the moduli space ${\cal M}_N$, the fields can be 
written as: 
\bee
{\bf X}^i &=& U \cdot {\rm diag.}( X_1^i, X_2^i, \cdots, X_N^i)
\cdot U^{-1} \otimes {\bf I}_{2 \times 2}
\nonumber \\
{\bf S}^{\dot a} &=& U \cdot {\rm diag.} ( S^{\dot a}_1, \, S^{\dot
a}_2 , \cdots, \, S^{\dot a}_N)
\cdot U^{-1} \otimes  {\bf I}_{2 \times 2}
\label{diagonalized}
\eee
where $U \in$U(N)$\subset$SO(2N) is the same for all bosonic and
spinor coordinate matrices. The tensor product ${\bf I}_{2 \times 2}$
represents image strings. This means that, in the Higgs phase, the
residual gauge symmetry is $S_N \times [{\bf Z}_2]^N$, where $S_N$
denotes the Weyl group of U(N)$\subset$SO(2N) acting on the gauge
invariant elements for
a given representation and the $[{\bf Z}_2]^N$ denotes
the discrete gauge symmetry Eq.~(\ref{z2gaugesymmetry}) acting 
nontrivially on the twisted sector fermions in the fundamental 
representation. Together with the twisted sector fermions
$(\chi^{(P)}_I, \chi^{(A)}_I)$, the eigenvalues $(X^i_I, S^{\dot a}_I)$ 
$(I = 1, \cdots, N)$
describe light-cone Green-Schwarz worldsheet fields of
N independent heterotic strings. 

The Hilbert space ${\cal H}_N$ of the above $S_N$ orbifold sigma
model is decomposed into twisted sectors ${\cal H}_g$~\cite{dmvv}. 
Each twisted
sector is labelled by
the conjugacy classes $[g]$ of the $S_N$ symmetric orbifold group.
Within a given twisted sector, the physical states are those invariant
under the centralizer subgroup $C_g$ of $g \subset U(N) \subset
SO(2N)$. Denote this invariant
subspace by ${\hat {\cal H}}_g$. Then, the total $S_N$ orbifold
Hilbert space is given schematically by
\be
{\cal H}(S^N ({\bf R}^8 \otimes G))
= \bigoplus_{[g]} {\hat {\cal H}}_g.
\ee
For the $S_N$ symmetric group, the conjugacy classes $[g]$ are 
characterized by all partitions of $N$:
\be
\sum_{n \ge 1} n N_n = N
\ee
where 
$N_n$ denotes the multiplicity of the irreducible
cyclic permutation of $n$-elements
in the decomposition of $g$ as:
\be
[g] = (1)^{N_1} (2)^{N_2} \cdots = \prod_{n \ge 1} (n)^{N_n}.
\ee
Associated to the above decomposition of the conjugacy classes $[g]$
, each twisted sectors are 
decomposed into  the product of $N_n$-fold symmetric tensor products
of $n$-element Hilbert subspaces ${\cal H}_{(n)}$:
\be
{\cal H}_{N_n} = \bigotimes_{ n \ge 1} S^{N_n} {\cal H}_{(n)}
\ee
where
\be
S^N {\cal H} \equiv \Big( {\cal H} \otimes {\cal H} \otimes
\cdots \otimes {\cal H} \Big)^{S_N}.
\ee
The Hilbert subspace ${\cal H}_{(n)}$ denotes the ${\bf Z}_n$ 
invariant subspace of states
of a single heterotic string with winding number $n$ around ${\bf S}_1$.
Each of the winding number $n$ heterotic string can be represented by a 
$(8,0)$ supersymmetric heterotic sigma model whose $n$ distinct
worldsheet fields $(X^i_I, S^{\dot a}_I, \chi^{(P)}_I, \chi^{(A)}_I)$ 
satisfy twisted boundary conditions
\bee
X^i_I (\sigma + 2 \pi) &=& + \,\, [ \, V_{(\pm)} \cdot X^i (\sigma) \cdot 
V^{-1}_{\rm (\pm)} \, ]_I
\nonumber \\
S^{\dot a}_{I} (\sigma + 2 \pi) &=& + \,\, 
[\, V_{(\pm)} \cdot S^{\dot a} (\sigma)
\cdot V^{-1}_{(\pm)} \, ]_I 
\nonumber \\
\chi^{(P)}_I (\sigma + 2 \pi) & = & + \,\, [\,  V_{(\pm)} \cdot
 \chi^{(P)} (\sigma) \, ]_I
\nonumber \\
\chi^{(A)}_I (\sigma + 2 \pi) &=& -  \,\, [\, V_{(\pm)} \cdot
\chi^{(A)} (\sigma) \,]_I
\hskip0.5cm (I = 1, \cdots, n)
\label{period}
\eee
where $V_{(\pm)}$ denote the `t Hooft's shift operators combined with
nontrivial $[{\bf Z}_2]^n$ bundle
\be
V_{(+)} = \left( \begin{array}{ccccc}
 & 1 & & &  \\
 & & 1 & &   \\
 & & & \ddots &   \\
& & & & 1 \\
 +1 & & &  & \\
\end{array} \right)_{n \times n},
\hskip1cm
V_{(-)} = \left( \begin{array}{ccccc}
 & 1 & & &  \\
 & & 1 & &   \\
 & & & \ddots &   \\
& & & & 1 \\
 -1 & & &  & \\
\end{array} \right)_{n \times n}.
\ee 
The action of ${\bf Z}_n$ in the definition of ${\cal H}_{(n)}$ is most
clearly seen by patching $n$ fields into a single field
$(X^i (\sigma), S^{\dot a} (\sigma), \chi^{(P)}, \chi^{(A)} )_n$
defined over $\sigma \in [0, 2 \pi n]$. The single fields are defined
by
\bee
X^i (\sigma + 2 \pi (I-1) ) &\equiv& X^i_I (\sigma) \,\, , 
\nonumber \\
S^{\dot a} (\sigma + 2 \pi (I-1) )  &\equiv& S^{\dot a}_I (\sigma) \,\, ,
\nonumber \\
\chi^{(P)} (\sigma + 2 \pi (I-1) ) &\equiv& (+)^{I-1} \, 
\chi^{(P)}_I (\sigma) \,\, , 
\nonumber \\
\chi^{(A)} (\sigma + 2 \pi (I-1) ) &\equiv& (-)^{I-1} \,
\chi^{(A)}_I (\sigma) \,\, . \hskip1.5cm (I = 1, \cdots, n)
\eee  
Then the action of the cyclic group 
${\bf Z}_n$ generated by ordered shift of the 
$n$-element sets $(X^i_I, S^{\dot a}_I, \chi^{(P)}_I,
\chi^{(A)}_I)$ turns into the coordinate shift 
$\sigma \rightarrow \sigma + 2 \pi I$
acting on the newly introduced fields 
$(X^i (\sigma), S^{\dot a} (\sigma), \chi^{(P)} (\sigma), 
\chi^{(A)} (\sigma) )_n$ for $(I=1, \cdots, n)$. 
Physically, these newly introduced fields are interpreted as worldsheet 
degrees of freedom of ${\bf Z}_n$ twisted, $n$-times long strings.
These fields satisfy boundary conditions  
\bee
  X^i(\sigma + 2 \pi n) &=&  + \,\, X^i (\sigma),
\nonumber \\
            S^{\dot a} (\sigma + 2 \pi n) &=& + \, \,  
S^{\dot a} (\sigma),
\nonumber \\
            \chi^{(P)} (\sigma + 2 \pi n) &=& (+)^n \, \eta \, 
\chi^{(P)} (\sigma), 
\nonumber \\
            \chi^{(A)} (\sigma + 2 \pi n) &=& (-)^n \, \eta \,
\chi^{(A)} (\sigma),
\label{znaction}
\eee
where $\eta \equiv \pm$ denote the action of $[{\bf Z}_2]^n$
bundle.
Therefore, in the normalization that sets the string winding
number to unity for all $n$, we find that the $n$-twisted
strings have $1/n$ fractional oscillator quantum numbers. 
In addition, the twisted sector fermion fields give rise to
various charged sectors that generate the spacetime heterotic 
gauge group. Denote the boundary conditions of twisted sector
fermion fields as $(\chi^{(P)}, \chi^{(A)})$.
From Eq.~(\ref{znaction}), one finds that $(+,+)$ and $(-,-)$ sectors 
arise when $n$ is even. Similarly, when $n$ is odd, $ (+,-)$ and $(-,+)$ 
sectors arise. 

So far, we have constructed the Hilbert space for a finite $N$, 
corresponding to $N$ maximal number of heterotic strings~\footnote{
The conventional light--cone string field is then reproduced by
Fourier transforming Fock space wave functions with respect to
the string number}. The total $L_0$ operator of the
$S_N$ CFT is decomposed into
\be
L_0^{\rm total} = \sum_{\{n_i\}} {1 \over n_i} L_0^{(n_i)}
\ee
in a twisted sector defined by cyclic permutations of length $n_i$'s.
The ${\bf Z}_n$ invariant Hilbert subspace is then described by 
those excitations of a single string of length $n_i$ satisfying
\be
L_0^{\rm total} - {\overline L}_0^{\rm total}
= {\bf Z} \hskip0.75cm {\rm viz.} \hskip0.75cm
L_0^{(n_i)} - {\overline L}_0^{(n_i)} = n_i {\bf Z}.
\ee 
To obtain the full Fock space of the second-quantized heterotic 
string we take the large $N$ limit in the following manner~\cite{dvv}.
In the light-cone formulation, length of the string is directly 
proportional to the total longitudinal momentum (recall that we
have adopted the normalization that the $n$-string winding number is
unity, hence, $p_+^{\rm tot} \equiv N/R_{11} = 1$.). 
By taking Hilbert subspaces
${\cal H}_{(n)}$ for which $n$ is linearly proportional to $N
\rightarrow \infty$, only those heterotic
string states with {\sl non-vanishing} longitudinal momentum
$p_+$ are allowed:
\be
0 < \, p_+ = {n \over N} \le 1.
\ee
These long strings are the only surviving ones in the large N limit.
Correspondingly, since we have normalized the string winding number
to unity, string statets af finite massive levels constitute 
of very low-energy ${\cal O}(1/N)$ oscillator excitations. 

In the N$\rightarrow \infty$ 
limit prescribed above, the ${\bf Z}_n$ invariant
Hilbert subspace then consists only of individual string configurations
satisfying 
\be
L_0^{(n_i)} - {\overline L}_0^{(n_i)} = 0.
\ee
All other string configurations with nonzero on the right-hand side
become infinitely heavy in the large N limit, hence, are not relevant
for low-energy dynamics. 

It is now straightforward to identify the heterotic string degrees of
freedom and the required GSO projections thereof.  
From Eq.~(\ref{znaction}) we have seen that the twisted sector fermions
give rise to $(+,+)$ and $(-,-)$ sectors when $n$ is
even, and $(+,-)$ and $(-,+)$
sectors when $n$ is odd. Since difference of the string length by one unit
corresponds to difference of $p_+ = {\cal O}(\pm 1/N)$, hence, becomes
completely negligible in N$\rightarrow \infty$ limit, 
we define a single heterotic string of a given $p_+ = n/N$ via a 
direct sum of
the $(+,-)$ and $(-,+)$ sectors from $n$-twisted string together
with the
$(+,+)$ sector from $(n+1)$-twisted string and the $(-,-)$ sector 
from
$(n-1)$-twisted string. With this definition, we then 
find that, in the string winding number normalization adopted above,
 the vacuum energy associated
with $(+,-)$ and $(-,+)$ sectors are zero, while with $(+,+)$ and $(-,-)$
sectors are $+1$ and $-1$ respectively.
This fits perfectly to the known charged spectra of the perturbative
heterotic string. 
At the same time, the two independent GSO projection operators of the 
 heterotic string theory are identified as follows.   
The first one is the diagonal $\eta = \pm$ in Eq.~(\ref{znaction})
corresponding to $(-)^{P+A}$. The second one is the diagonal cyclic
permutation by $\pm 1$, the first factor in Eq.~(\ref{znaction}). This 
gives rise to the projection via anti-periodic boundary condition,
 $(-)^A$. We therefore obtain two independent GSO 
projection operators,
$(-)^P$ and $(-)^A$, needed to reproduce the $E_8 \times E_8$ 
heterotic string. 

We should emphasize again the importance of N$\rightarrow
\infty$ limit in identifying these two independent GSO projection 
operators and the resulting charged spectrum thereof.
While the first GSO operator $(-)^{P+A}$ relevant for 
SO(32) heterotic string is well defined even for finite N, the 
second one $(-)^A$ arises by construction only in the
large N limit. Related to this, while the $(+,-)$ and $(-,+)$ 
sectors are present for $n$ odd as explained above, for finite
N they are projected out by the $(-)^{P+A}$ GSO operator completely.
We conclude that only SO(32) heterotic strings are present for finite 
N heterotic M(atrix) string theory. 

%%%%%%%%%%%%%%%%%%%%%%%%%%%%%%%%%%%%%%%%%%%%%%%%%%%%%%%%%%%%%%%%%%%%%%%%%%%
\subsection{Perturbative Interactions of Heterotic M(atrix) String}
%%%%%%%%%%%%%%%%%%%%%%%%%%%%%%%%%%%%%%%%%%%%%%%%%%%%%%%%%%%%%%%%%%%%%%%%%%%

So far, we have shown that the strict infrared limit of the $(8,0)$
supersymmetric chiral gauge theory gives rise to free heterotic string
in the light-cone Green-Schwarz form. We have seen that, in this limit,
each twisted sector of the $S^N$ orbifold is a Fock space of a
definite string number and longitudinal momentum. 
For free heterotic string,
each Fock space of multi-heterotic string states spans 
a superselection sector.

In this section, we consider the effect of non-zero but small heterotic
string coupling constant. Recall that the height of potential barrier
of ${\bf X}^i$ for non-abelian symmetry restoration is proportional 
to $g_H^{-2}$. Large-N counting shows that the least barrier, hence,
easiest restoration is for SO(4) $\subset$ SO(2N) corresponding to two 
overlapping heterotic strings. Once they overlap, SO(4) gauge field
fluctuations induce a non-zero intercommuting transition amplitude
between the two heterotic strings. Indeed, this is precisely the 
elementary joining and splitting interactions between two strings
since the string number between in- and out-string states differ by one.

It is also possible to represent the elementary string interaction
in terms of perturbed $S^N$ orbifold conformal field theory.
The perturbation is described by a {\sl local} operator 
since,
being induced by non-abelian gauge field fluctuations in the Higgs 
phase, the transition is a {\sl local} process~\footnote{The characteristic
length scale of the massive charged gauge boson is $\sim
g_H / \langle X_{IJ} \rangle$, where $X_{IJ}$ denote separation distance
between I-th and J-th strings.} on the worldsheet 
$(\sigma, \tau)$~\footnote{as well as being local in spacetime.}.
A candidate local perturbation operator should satisfy several 
physical conditions.
First, since the operator represents elementary string joining and
splitting processes, it should correspond to an appropriate twist
operator that intercommutes two overlapping heterotic strings. 
Second, the operator should
be a singlet under $(8,0)$ supersymmetry, $SO(9,1)$ Lorentz transformation
and $SO(16) \otimes SO(16)$ Kac-Moody symmetry. Third, the operator 
should be an irrelevant operator to guarantee that the unperturbed $S^N$ 
conformal field theory is an infrared stable fixed point. Fourth,
the leading irrelevant operator should have scaling dimension 3 so
that the perturbation coupling parameter corresponds precisely to 
the heterotic string coupling constant itself. 

For the Type IIA M(atrix) string theory, DVV have shown the existence 
of a local perturbation operator satisfying all of these requirements. 
The corresponding $S^N$ orbifold conformal field theory has $(8,8)$
supersymmetries. In contrast, the heterotic M(atrix) string theory 
corresponds to a $S^N$ conformal field theory with chiral 
$(8,0)$ supersymmetry~\cite{brinketal}
 and $SO(16) \times SO(16)$ Kac-Moody 
symmetry. As such, it is not a priori obvious that 
 there again exists a local perturbation operator satisfying all of
the above requirements. We now show that such an operator
of scaling dimension $(3/2, 3/2)$ indeed exists so that the
perturbed $S^N$ conformal field theory is described by:
\bee
S_{\rm (8,0)~ int} &=&
 S_{(8,0)~ \rm free} + g_H \int {d^2 z \over \ell_s^2}
\, [\ell_s^3 \, {\cal O}_{({3\over 2}, {3 \over 2})}] + \cdots
,
\nonumber \\
&& {\cal O}_{({3 \over 2}, {3 \over 2})}  (z, {\overline z})
= {\cal O}_{( {3 \over 2}, 0)} (z) \cdot
{\cal O}_{(0, {3 \over 2})} ({\overline z}),
\label{schematic}
\eee
where ellipses denote higher-order contact interactions.
We now derive the operator as a tensor product of left-moving
supersymmetric descendent twist operator of dimension 3/2 and 
right-moving bosonic twist operator of dimension 3/2.

Consider first the left-moving supersymmetric sector. This sector
gives rise to all of the ${\cal N}= 16$ spacetime supersymmetry 
generators in the light-cone Green-Schwarz formulation:
\be
{\bf Q}^{\dot a} = {1 \over \sqrt N} \oint d \sigma \, 
\sum_{I=1}^N S^{\dot a}_I
, \hskip1.3cm
{\bf Q}^{ a} = {\sqrt N} \oint d \sigma \, \sum_{I=1}^N
\sigma^i_{a \dot a } S^{\dot a}_I \partial_z X^i_I.
\ee

As we have discussed, the leading order perturbation is given by
intercommuting interactions between the two free
heterotic strings, say, I-th and J-th. Each of them is described
by the unperturbed conformal field theory. 
The interaction corresponds to ${\bf Z}_2$ twist operator associated
with a discrete gauge symmetry ${\bf Z}_2 \in SU(2) \subset SO(4)$
acting on a product of I-th and J-th string conformal field theories.  
Decompose the product into a conformal field theory of center-of-mass 
motion and the another of relative motion. The ${\bf Z}_2$ acts on
the conformal fields of relative motion
$X_{IJ}^i \equiv (X_I^i - X_J^i), \,\, S^{\dot a}_{IJ} \equiv 
(S^{\dot a}_I - S^{\dot a}_J)$:
\bee
{\bf Z}_2 : X_{IJ}^i &\leftrightarrow& - X^i_{IJ} 
\nonumber \\
            S^{\dot a}_{IJ} & \leftrightarrow & - S^{\dot a}_{IJ}.
\label{z2action}
\eee
This is exactly the same as either chiral sector of Type IIA M(atrix)
string theory~\cite{dvv} and corresponds to the well-known supersymmetric 
${\bf R}^8/{\bf Z}_2$ orbifold. The twisted sector of the orbifold arises
from ${X}^i_{IJ} = 0$, viz. when the two heterotic strings overlap. 

Twist field of the above supersymmetric orbifold is a product of
bosonic and fermions parts. Associated to $X^i_{IJ}$ is the bosonic twist 
operator $\sigma_{IJ} $ defined through the operator product expansion:
\bee
\partial_z X^i_{IJ} (z) \cdot \sigma_{IJ} (w) 
&\sim& {1 \over (z - w)^{1 \over 2}} \, \tau^i_{IJ} (w),
\nonumber \\
\partial_z X^i_{IJ} (z) \cdot \tau^i_{IJ}(\omega)
&\sim& {1 \over 2} \delta^{ij} {1 \over (z - {\omega})^{3 \over 2}} \, 
\sigma_{IJ}(\omega)
+ 2 \delta^{ij} {1 \over (z - \omega)^{1 \over 2} } \partial_\omega
\sigma_{IJ} (\omega).
\eee
Here, $\tau^i$ denotes the excited twist 
field transforming as ${\bf 8}_{\rm v}$ 
under Spin(8).
Associated to $S^{\dot a}_{IJ}$ are a pair of spin fields 
$(\Sigma^i_{IJ}, \Sigma^{a}_{IJ})$ defined through the 
operator product expansions
\bee
S^{\dot a}_{IJ}(z) \cdot \Sigma^i_{IJ} (\omega) &\sim& {1 \over
(z-\omega)^{{1 \over 2}}}
\, \sigma^i_{{\dot a} a } \Sigma^{ a}_{IJ} (\omega),
\nonumber \\
S^{\dot a}_{IJ} (z) \cdot \Sigma^{ a}_{IJ} (\omega)
 &\sim& {1 \over (z - \omega)^{{1 \over 2}}} \, \sigma^i_{\dot a a} 
\Sigma^i_{IJ} (\omega).
\eee
Under Spin(8), the spin fields $(\Sigma^i, \Sigma^{ a})$
transform as $({\bf 8}_{\rm v}, {\bf 8}_{\rm s})$. 
Taking operator product expansions with the energy-momentum tensor
\be
T_{IJ} (z) = {1 \over 2} \partial_z X^i_{IJ} \partial_z X^i_{IJ}
+ {1 \over 2} S^{\dot a}_{IJ} \partial_z S^{\dot a}_{IJ},
\ee
we find the conformal dimensions of the twist operators 
as $[\sigma] = 1/2, \,\,\, [\tau_i] = 1, \,\,\,  
[ \Sigma^i]= [ \Sigma^{ a}] = 1/ 2$ respectively. 
The unique $(8,0)$ supersymmetric and $SO(9,1)$ Lorentz invariant
operator is the supersymmetry descendent of one of the chiral primary operators
$\sigma \, \Sigma^{ a}$: 
\bee
{\cal O}_{({3 \over 2}, 0)} (0) &=& \oint_{z=0} d z \, 
\Big( \sigma^i_{ \dot a a } S^{\dot a} \partial_z X^i \Big)_{IJ} (z) 
\Big(\sigma \Sigma^{ a}\Big)_{IJ} (0)
\nonumber \\
&=& \oint_{z=0} {d z \over z^{1 \over 2}} \,
\Big( \partial_z X^i \Sigma^i \Big)_{IJ} (z) \, \sigma_{IJ}(0)
\nonumber \\
&=& \Big(\tau^i \Sigma^i \Big)_{IJ} (0).
\label{opinsertion}
\eee

Next, consider the right-moving bosonic sector. The sector consists of
${X}^i$'s and thirty-two Majorana fermions $\chi^{(P)}, \, 
\chi^{(A)}$. It is convenient to bosonize the
chiral fermions into compact chiral bosons 
$Y^A, \,\,\, (A = 1, \cdots, 16)$.
We again analyze the action of ${\bf Z}_2$ discrete gauge group 
$\subset SO(4)$ to
the right-moving sector when I-th and J-th heterotic string overlap.   
Again, decomposing the product of two anti-holomorphic 
conformal field theories into
center-of-mass and relative motion conformal field theories, 
the ${\bf Z}_2$ twist acts on the chiral bosons as:
\bee
{\bf Z}_2 : \partial_{\overline z} X^i_{IJ} 
&\rightarrow& - \partial_{\overline z} X^i_{IJ} ,
\nonumber \\
\partial_{\overline z} Y^A_{IJ} &\rightarrow&
- \partial_{\overline z} Y^i_{IJ}.
\label{z2actionright}
\eee
Hence, the relative-motion conformal field theory corresponds to 
$({\bf R}^8 \times {\bf T}^{16}) / {\bf Z}_2$ orbifold.
For our purposes, however, it is sufficient to take the $Y^A_{IJ}
\in {\bf R}$. This is because $Y^A_{IJ} \approx - Y^A_{IJ} + 2 \pi$
is equivalent to $Y^A_I \rightarrow (Y^A_J +  \pi)$, $Y^A_J \rightarrow
(Y^A_I - \pi)$, which is part of discrete gauge symmetries in the
heterotic string~\cite{dineetal}, and because 
the intercommuting I-th and J-th heterotic
string is a purely {\sl local} process in $Y^A_{IJ}$ space as well as
in spacetime. Therefore, the left-moving sector is given by the 
well-known ${\overline c}=24$
${\bf R}^{24} / {\bf Z}_2$ orbifold conformal field 
theory~\cite{dixonetal}. 
The twist operator is again defined through the operator product
expansions:
\bee
\partial_{\overline z} X^i_{IJ} ({\overline z}) 
\cdot {\overline \sigma}_{IJ} ({\overline \omega})
&\sim& {1 \over ({\overline z} - {\overline \omega})^{1 \over 2}} 
\, {\overline \tau}^i ({\overline \omega}),
\nonumber \\
\partial_{\overline z} Y^A_{IJ} ({\overline z})
\cdot {\overline \sigma}_{IJ} ({\overline \omega})  
&\sim& {1 \over ({\overline z} - {\overline \omega})^{1 \over 2}} 
\, {\overline \tau}^A ({\overline \omega}).
\eee 
The energy-momentum tensor is given by
\be
{\overline T}({\overline z}) 
= {1 \over 2} \partial_{\overline z} X^i \partial_{\overline z} X^i
+ {1 \over 2} \partial_{\overline z} Y^A \partial_{\overline z} Y^A.
\ee
From this, we find that the twist operator ${\overline \sigma}$ 
has conformal
dimension $24/16 = 3/2$, while the excited twist operators 
${\overline \tau}^i$ have
conformal dimension 2. We thus find a unique
leading irrelevant twist operator on the left-moving sector as:
\be
{\overline {\cal O}}_{(0, {3 \over 2})} (0) = {\overline \sigma}( 0).
\ee
It is straightforward to see that the above operator
is unique. To see this, consider the anti-holomorphic partition function 
${\cal Z} ({\overline q})$ 
of the second symmetric product of the heterotic string theory.
We will explicitly label the boundary conditions along $(\sigma, \tau)$
directions as ${\cal Z}_{(\sigma, \tau)}$.
Denote a single heterotic string partition function as $Z(
{\overline q}) = {\overline q}^{-1} + 24 + {\cal O}({\overline q}) $.
Then, in the $(+,+)$ sector, the second-symmetric product 
 partition function is given by
${\cal Z}_{(+,+)}({\overline q}) = (Z({\overline q}))^2 = {\overline q}^{-2} 
+ \cdots$. On the other hand, the partition function in the
$(+,-)$ is given by ${\cal Z}_{(+,-)}({\overline q}) 
= Z({\overline q}^2)$.
By modular invariance, the twisted sector $(-,+)$
then gives rise to the second symmetric product partition function 
${\cal Z}_{(-,+)} ({\overline q}) = 
Z({\overline q}^{1\over 2})$.
Therefore, in the second symmetric space, ${\overline c}/24 = 2$ and
we find the twisted sector spectrum as:
\be
{\cal Z}_{(-,+)} ({\overline q}) = 
{\rm Tr}_{(-,+)} {\overline q}^{({\overline {\rm L}}_0 - {{\overline c} 
\over 24})}
= Z({\overline q}^{1 \over 2} ) = 1 \cdot {\overline q}^{-{1
\over 2}} + \cdots.
\ee
It shows that the ground-state of the twisted sector has multiplicity
one and is created by the afore-mentioned conformal dimension $(0, 3/2)$
twist operator $\overline \sigma$ of the ${\overline c} = 24$
${\bf R}^{24}/{\bf Z}_2$ orbifold conformal field theory.

The complete form of the cubic string interaction operator 
is obtained by tensoring the left-moving and the right-moving 
twist operators for every pairs of $(IJ)$. This yields
\bee
{\cal O}_{({3 \over 2}, {3 \over 2})} (0)
= {\cal O}_{({3 \over 2}, 0)} (0) \cdot {\overline {\cal O}}_{(0, 
{3 \over 2})} (0)
&=&
\sum_{I < J}
\oint_{z=0}
d z 
 \, \Big(\sigma^i_{\dot a a}
S^{\dot a} \partial_z X^i \Big)_{IJ} (z) 
\Big( \Sigma^a \sigma \cdot {\overline \sigma} \Big)_{IJ} (0)
\label{gsform} \\
&=&
\sum_{I < J}
\oint_{z = 0} {d z \over z^{1 \over 2}}
\, \Big( \partial_z X^i \Sigma^i \Big)_{IJ}
(z) \, \Big( \sigma \cdot {\overline \sigma} \Big)_{IJ} (0).
\label{nsform}
\eee
The operator has a scaling dimension $(3/2, 3/2)$. As such, the
corresponding worldsheet coupling constant scales linearly with $g_H$,
a result which asserts that this operators should be identified with
three string joining and splitting interactions. In the next section, we 
show explicitly that this is indeed the case by comparing with the
heterotic string light-cone worldsheets in Mandelstam's approach.
Given that the structure and field content of the conformal field theory
are so different from those of type II string, the result may be viewed
as a nontrivial consistency check to the heterotic M(atrix) string theory.

%%%%%%%%%%%%%%%%%%%%%%%%%%%%%%%%%%%%%%%%%%%%%%%%%%%%%%%%%%%%%%%%%%%%%%%%%
\subsection{Comparison with Conventional Light-Cone Heterotic String}
%%%%%%%%%%%%%%%%%%%%%%%%%%%%%%%%%%%%%%%%%%%%%%%%%%%%%%%%%%%%%%%%%%%%%%%%%
Having determined the conformal field theory and the leading irrelevant
operator corresponding to joining and splitting
interactions of heterotic M(atrix) string, 
it is of interest to compare them with the conventional light-cone 
heterotic string field theory and cubic interactions thereof. 
Below we show that the leading irrelevant operator of dimension (3/2,
3/2) in the heterotic
M(atrix) string theory is precisely the cubic interaction vertex
of light-cone heterotic string both in Green-Schwarz and 
Neveu-Schwarz-Ramond formulations.

The conventional light-cone Green-Schwarz heterotic string is described
in terms of the following worldsheet fields: bosonic coordinates
$X^i(\sigma, \tau)$, fermionic real coordinates 
$S^{\dot a} (\sigma, \tau)$ and 32 fermionic real coordinates
$\chi^A(\sigma, \tau)$ parametrizing sixteen--dimensional
even self-dual lattice of either $E_8 \times E_8$ or Spin$(32)/Z_2$
gauge groups.
They have $SO(8)$ R--symmetry quantum numbers
$({\bf 8}_{\rm v}, {\bf 8}_{\rm c}, {\bf 1})$ respectively.
The light-cone gauge Lagrangian of a free heterotic string is
\be
L_{\rm free} = \int d \sigma \,
\Big( \partial_\rho X^i \partial_{\overline \rho} X^i
+ i S^{\dot a} \partial_{\overline \rho} S^{\dot a} + 
i \chi^A \partial_\rho \chi^A \Big)
\label{gsfreeaction}
\ee
where $\rho \equiv \tau + \sigma$ and ${\overline \rho} = \tau - \sigma$.
We will freely use the same notation for Euclidean worldsheet
$\rho \equiv \tau + i \sigma, {\overline \rho} \equiv \tau - i \sigma$,
for which the fermionic fields $S^{\dot a}, \chi^A$ and their Hermitian
conjugates should be interpreted  as independent complex-valued fields.
Global spacetime supersymmetry requires that both $X^i, S^{\dot a}$ 
satisfy
periodic boundary conditions. The 32 fermionic fields $\chi^A$ are
SO(8) singlets, hence, can take either periodic or anti-periodic
boundary conditions subject to compatibility with GSO projections. 

Using SO(8) triality, it can be shown~\cite{wittenequiv} that 
the free heterotic string action 
Eq.~(\ref{gsfreeaction}) in light-cone Green-Schwarz formulation
 is equivalent
to the free heterotic string action in light-cone Neveu-Schwarz-Ramond
formulation. This is true in so far as both periodic and anti-periodic
boundary 
conditions and accompanying GSO projections are assumed for the latter
formulation. The
light-cone worldsheet fermions $\Psi^i \,\, (i=1, \cdots, 8)$
in the 
Neveu-Schwarz-Ramond formulation can be identified with the
conformal dimension (1/2, 0) ${\bf 8}_{\rm v}$ spin field $\Sigma^i$.

The light-cone cubic interaction vertex takes a simple form in
the Neveu-Schwarz-Ramond formulation. In this formulation, 
Mandelstam~\cite{mandelstamnsr} has found that the light-cone worldsheet
Lagrangian associated with the joining and splitting cubic interaction 
at $\rho = {\tilde \rho}$ is given by
\be
L_{\rm int}^{\rm (NSR)}({\tilde \rho}) 
= \lambda_H
\int d \sigma \, {\cal O}_{\rm NSR} (\rho, {\tilde \rho})
\prod_\sigma \delta (X^i_{\rm in} - X^i_{\rm out})
\delta (\Psi^i_{\rm in} - \Psi^i_{\rm out})
\delta ( \chi^A_{\rm in} - \chi^A_{\rm out})
\label{nsrvertex}
\ee
where the interaction location--dependent operator insertion is
\be
{\cal O}_{\rm NSR} (\rho, {\tilde \rho})
 = {\rm Lim}_{\rho \rightarrow {\tilde \rho}}
(\rho - {\tilde \rho})^{3 \over 4} \,
\Psi^i \partial_\rho X^i.
\label{nsoperator}
\ee
Here, $\lambda_H$ denotes the cubic interaction coupling parameter
and the overlap delta functional is between the initial and the 
final heterotic string wave functionals around the interaction point.
The operator insertion Eq.~(\ref{nsoperator}) is only for the
right-moving, supersymmetric sector. 
On the other hand, the overlap delta functional in 
Eq.~(\ref{nsrvertex}) acts on both left- and right-moving sectors,
much the same way as in the bosonic string.   

The light-cone cubic interaction of the Green-Schwarz heterotic string 
has been obtained by Green and Schwarz~\cite{greenschwarz} and by
Mandelstam~\cite{mandelstamgs}. It
is most conveniently described in terms of $SU(4) \times U(1) \subset
SO(8)$. Conventionally the bosonic and fermionic worldsheet fields are
decomposed into
irreducible representations of $SU(4) \times U(1)$, 
\bee
&& {\bf 8}_{\rm v} 
= \Big(X^R \equiv {1 \over \sqrt 2} (X^1 + i X^2), \hskip0.2cm
X^L \equiv {1 \over \sqrt 2} (X^1 - i X^2), \hskip0.2cm
Y^{i \ge 3} \equiv X^{i \ge 3} \Big) = ({\bf 1}_{+1}, {\bf 6}_0, 
{\bf 1}_{-1}) ,
\nonumber \\
&& {\bf 8}_{\rm c}
= \, \Big( {\tilde S}^{ a} \equiv {1 \over \sqrt 2} 
(S^{\dot a}  + i S^{\dot a + 4} ), \,\,\,
{\tilde S}^{\dagger a} \equiv 
{1 \over \sqrt 2} (S^{\dot a} - i S^{\dot a + 4})\Big)
\,\, = \,\, ( {\bf 4}_{-{1 \over 2}}, \,\,\,
{\overline {\bf 4}}_{+{1 \over 2}}) .
\eee
In the Green-Schwarz formulation, the physical ground-state form
two inequivalent eight-fold degenerate multiplets that are created 
by acting the zero-modes of ${\tilde S}^{ a}$. Physically they 
represent ground state spanned by ${\bf 8}_{\rm v}$ 
massless vector boson whose polarization is
either R = (1 + i 2) = ${\bf 1}_+$ or L = (1 -- i 2) = ${\bf 1}_-$ type.
With these vacuum choices light-cone worldsheet Lagrangian describing 
the cubic interaction at $\rho = {\tilde \rho}$ is given 
by~\cite{mandelstamgs}:
\be
L_{\rm int}^{\rm (GS)} ({\tilde \rho})=
\lambda_H \int d \sigma \,
 {\cal O}_{\rm GS}^{\pm} (\rho, {\overline \rho}) \, 
\prod_\sigma \delta (X^i_{\rm in} - X^i_{\rm out})
\, \delta (S^a_{\rm in} - S^a_{\rm out}) \,
\delta (\chi^A_{\rm in} - \chi^A_{\rm out}),
\label{gsvertex}
\ee
where 
\be
{\cal O}_{\rm GS}^+
= {\rm Lim}_{\rho \rightarrow {\tilde \rho}} 
( \rho - {\tilde \rho})^{1 \over 2} \,
\Big[
\partial_\rho X^L + {1 \over 2} (\rho - {\tilde \rho}) \,
\lambda^i_{{\dot a} {\dot b}}
\partial_\rho Y^i {\tilde S}^{\dagger \dot a} {\tilde S}^{\dagger \dot b}
+ (\rho - {\tilde \rho})^2 \, \partial_\rho X^R {\tilde S}^{1 \dagger}
\cdots {\tilde S}^{4 \dagger} \Big]
\label{gsvertex+}
\ee
for L = (1 + i2) =${\bf 1}_+$ (Mandelstam's `empty ground-state')
boundary condition
and
\be
{\cal O}_{\rm GS}^-
= {\rm Lim}_{\rho \rightarrow {\tilde \rho}} 
( \rho - {\tilde \rho})^{1 \over 2} \,
\Big[ 
\partial_\rho X^R + {1 \over 2} (\rho - {\tilde \rho}) \,
\lambda^i_{{\dot a} {\dot b}}
\partial_\rho Y^i {\tilde S}^{\dot a} {\tilde S}^{\dot b}
+ (\rho - {\tilde \rho})^2 \, \partial_\rho X^L {\tilde S}^1
\cdots {\tilde S}^4 \Big]
\label{gsvertex-}
\ee
for R = (1 -- i 2) =${\bf 1}_-$ (Mandelstam's `full ground-state')
boundary condition respectively.

The above choice of physical ground-state is by no means unique.
Indeed, one can choose the ground state spanned by massless 
spinor ${\bf 8}_{\rm s}$ instead of massless vector ${\bf 8}_{\rm v}$.
In this case,  the SU(4)$\times$U(1) decompositions are
${\bf 8}_{\rm s} = ({\bf 1}_{+1}, {\bf 6}_0, {\bf 1}_{-1})$,
while ${\bf 8}_{\rm v} = ({\bf 4}_{-{1 \over 2}}, {\overline {\bf 4}}_{+{1
\over 2}})$, viz. $X^i = (Z^a \equiv X^a - i X^{a + 4}, Z^{\dagger a} \equiv
X^a + i X^{a + 4} )$. Correspondingly, the required operator insertions are
given by
\be
{\cal O}_{\rm GS}^+ 
= {\rm Lim}_{\rho \rightarrow {\tilde \rho}}
\Big[ (\rho - {\tilde \rho}) {\tilde S}^a
 \partial_\rho Z^a + (\rho - {\tilde \rho})^2
\partial_\rho Z^1 \, {\tilde S}^2 \, {\tilde S}^3 \, {\tilde S}^4
+ ({\rm perm.}) \Big]
\label{gsferm+}
\ee
for fermionic L = (1 + i 2) = ${\bf 1}_+$ boundary condition
and 
\be
{\cal O}_{\rm GS}^-
= {\rm Lim}_{\rho \rightarrow {\tilde \rho}}
\Big[ (\rho - {\tilde \rho}) {\tilde S}^{\dagger a} 
\partial_\rho Z^a  + (\rho - {\tilde \rho})^2
\partial_\rho Z^1 {\tilde S}^{\dagger 2} {\tilde S}^{\dagger 3}
{\tilde S}^{\dagger 4} + ({\rm perm.}) \Big]
\label{gsferm-}
\ee
for fermionic R = (1 -- i 2) = ${\bf 1}_-$ boundary condition
respectively.

The nonlinear spinor terms in Eqs.~(\ref{gsferm+},~\ref{gsferm-})
can be linearized further~\cite{berkovits}. 
Instead of integrating over worldsheets
with a fixed set of interaction-point boundary conditions, integrating
over both types of boundary conditions at the interaction points 
turns cubic in ${\tilde S}^a, {\tilde S}^{\dagger a}$ into 
linear ${\tilde S}^{\dagger a}, {\tilde S}^a$ respectively. 
Note that we have still kept fixed the boundary conditions at the 
external string endpoints. With these proviso, the operator insertion
for the Green-Schwarz cubic interaction turns into a single 
operator:
\be
{\cal O}_{\rm GS} \equiv {\cal O}^+_{\rm GS} + {\cal O}^-_{\rm GS} = 
{\rm Lim}_{\rho \rightarrow {\tilde \rho}}
(\rho - {\tilde \rho}) \Big[ {\tilde S}^a \partial_\rho Z^{\dagger a}
+  {\tilde S}^{\dagger a} \partial_\rho Z^a \Big].
\label{gsoperator}
\ee

It is now straightforward to compare the cubic interaction 
worldsheet Lagrangian Eqs.~(\ref{nsrvertex},~\ref{gsvertex})
of conventional light-cone heterotic string
and the unique leading irrelevant operator
Eqs.~(\ref{gsform},~\ref{nsform}). First, near the interaction
points, the light-cone worldsheet coordinates $(\rho, {\overline \rho})$
is a double--cover of the conformal
field theory local coordinates $(z, {\overline z})$.
Once conformal mapping is made to the latter coordinates, the
factors $(\rho - {\tilde \rho})^{3/4} \partial_\rho$ in 
Eq.~(\ref{nsoperator}) and $(\rho - {\tilde \rho}) \partial_\rho$ 
in Eq.~(\ref{gsoperator}) are mapped into $z^{-{1/2}} \partial_z$
in Eq.~(\ref{nsform}) and $\partial_z$ in Eq.~(\ref{gsform}) respectively.
Second, in order to represent the overlap delta functionals
in Eqs.~(\ref{gsvertex},~\ref{nsrvertex}), it is necessary to arrange the
second Riemann sheet at each interaction point to cover the
two joining or splitting strings. In the $(8,0)$
conformal field theory, this is achieved precisely by inserting 
${\bf Z}_2$ bosonic twisted operator $\sigma \cdot {\overline \sigma}$. 
In addition, in the Green-Schwarz formulation with fermionic boundary
conditions, the ${\bf 8}_{\rm s}$
spin field $\Sigma^a$ has to be inserted to allow
both types of boundary conditions in Eq.~(\ref{gsoperator}).
Finally, the conventional insertion operators Eqs.~(\ref{nsoperator},
~\ref{gsoperator}) are identified with the holomorphic operators
$\Sigma^i \partial_z X^i$ and $\sigma^I_{\dot a a} S^{\dot a}
\partial_z X^i$ in Eqs.~(\ref{gsform},~\ref{nsform}) respectively.
Putting these correspondences together we find that the 
leading irrelevant operator of $(8,0)$ superconformal field
theory indeed matches with the light-cone cubic interaction vertices of
heterotic strings
either in Green-Schwarz or in Neveu-Schwarz-Ramond formulations.

In the conventional light-cone interacting string picture, it has been
known that higher-order contact terms between two or more incoming
and outcoming string sets have to be introduced~\cite{contactterm}
in order to ensure a stable ground state. 
These contact terms
cancel divergences caused by two colliding cubic interaction vertices 
on the light-cone worldsheet. Since exact form of these contact terms
either in the Green-Schwarz or in the Neveu-Schwarz-Ramond formulations
are not known presently, we will not make any attempt to compare 
them with candidate sub-leading irrelevant operators in the $S_N$
conformal field theory.

%%%%%%%%%%%%%%%%%%%%%%%%%%%%%%%%%%%%%%%%%%%%%%%%%%%%%%%%%%%%%%%%%%%%%%%%%%%
\section{Discussions}
%%%%%%%%%%%%%%%%%%%%%%%%%%%%%%%%%%%%%%%%%%%%%%%%%%%%%%%%%%%%%%%%%%%%%%%%%%%
In this paper, we have extended the DVV proposal to the heterotic M(atrix)
theory. The resulting heterotic M(atrix) {\sl string} theory, which provides
 with a non-perturbative description of second-quantized heterotic strings,
is defined by $(8,0)$ supersymmetric chiral gauge theory with gauge group
SO(2N) in large N limit. We have checked that the theory is consistent with
known properties of conventional heterotic string. The leading irrelevant
operator of dimension 3 in the strong coupling expansion agrees with the
joining and splitting cubic interaction vertices of light-cone heterotic
string either in Green-Schwarz or in Neveu-Schwarz-Ramond formulation.

We would like to conclude with a highly speculative remark on a possible
M(atrix) theory description of bosonic strings.
It is well-known that bosonic Yang-Mills theory in twenty-six dimensions is
rather special~\cite{nepomechie}. The regularized one-loop effective action of
$d$-dimensional Yang-Mills theory is given by
\bee
\Gamma_{\rm d} =
&-& {\rm Tr} \int {d^d x \over (4 \pi)^{d/2}} \,
\Big[ {2 - d \over 2} {\Lambda^d \over d}
+ {26 - d \over 24} {\Lambda^{(d-4)} \over (d - 4)}
             F_{MN}^2 \nonumber \\
&-& {\Lambda^{(d-6)} \over (d-6)}
\Big( {(42 - d) \over 120} (D_M F_{MN})^2 +
{(2 - d) \over 144} F^3_{MN} \Big) + \cdots \Big].
\eee
For d=26, the gauge kinetic term does not receive radiative correction
at all, a feature shared by the ten-dimensional super-Yang-Mills theory.
We expect that this non-renormalization remains the same even after
dimensional reductions. For example, the four-dimensional Yang-Mills theory
with 22 pseudo-scalar fields in the adjoint representation has a vanishing
beta function at one loop and the renormalization group infrared fixed points for the scalar quartic couplings $\lambda_1 {\rm Tr} (X^i X^j X^i X^j)
$ and $\lambda_2 {\rm Tr} (X^i X^j X^j X^i)$ at
$\lambda_1 = \lambda_2 = g^2/(2 \pm {\sqrt {8/3}})$. 
Given this non-renormalization feature unique
to twenty-six dimensional Yang-Mills theory, one may wonder if it is possible
to construct M(atrix) string theory following DVV proposal for bosonic string
as well despite the absence of supersymmetry and BPS states.

The bosonic strings also have D-brane extended solitons ($0 \le p \le 25$)
whose tension scales as $1/g_B$ for weak string coupling $g_B \ll 1$.
Given the observation that the leading order string effective action of
graviton, dilaton and antisymmetric tensor field may be derived from an
Einstein gravity in $d=27$, let us make an 
assumption that the 27-th
`quantum' dimension decompactifies as the string coupling $g_B$ becomes large.
For D0-brane, the dilaton exchange force may be interpreted as the
27-th diagonal component of $d=27$ metric. Gravi-photon is suppressed
by compactifying 27-th direction on an orbifold rather than on a circle.
Likewise, its mass may be interpreted as 27-th Kaluza-Klein momentum of 
a massless excitation in $d=27$. 
In the infinite boost limit, the light-front
view of a bosonic string is that infinitely many D0-branes are threaded
densely on the bosonic string. This hints that D0-branes and Yang-Mills
gauge fields gluing them are the fundamental partons, the same content as the
strongly coupled superstrings. This should not be surprising since the infinite
momentum boost kinematics has little to do with supersymmetry.

Given the above observation,
it is quite possible that large N limit of
$(1+1)$-dimensional $U(N)$ gauge theory with 24 adjoint matter fields
${\bf X}^i \,\, (i=1,\cdots, 24)$ describes second-quantized bosonic strings
in light-cone formulation. As a variant of the DVV proposal, this may be
taken as a definition of bosonic M(atrix) string theory. The Yang-Mills
coupling $g_{\rm YM}$
scales with the bosonic string coupling $g_B$ the same way as superstring
cases: $g_{\rm YM}^{-2} = \ell^2_s g^2_{\rm B}$. If the strong coupling
limit of the Yang-Mills theory is confining and flows to a nontrivial fixed
point with manifest $SO(24)$ rotational symmetry, the bosonic M(atrix)
string theory is locally described by a $(c, {\overline c}) = (24, 24) $
conformal field theory defined on a symmetric product space orbifold
$S^N {\bf R}^{24} \equiv ({\bf R}^{24})^N/S_N$. Because the theory is
confining, only entries of diagonalized ${\bf X}^i$'s are observables.
Since each of the left- and the right-moving sectors are the same copies
as ${\bf R}^{24}/{\bf Z}_2$ orbifold conformal field theory sector of
the heterotic string, the leading irrelevant operator with manifest
SO(24) rotational symmetry is identified with ${\cal O}_{\rm B}
= \sigma \cdot {\overline \sigma}$. This operator has scaling dimension
$(3/2, \, 3/2)$, hence, worldsheet coupling associated with the
perturbation is proportional linearly to $g_B$, much the same as
the superstring cases. Furthermore, in Section 3.3, 
we have shown that perturbation by the operator
$\sigma \cdot {\overline \sigma}$ is equivalent to arranging the second
Riemann-sheet at each interaction points so that joining and splitting
string wave funcationals do overlap. The overlap delta functional is
all one needs to describe the joining and splitting cubic interactions
in light-cone bosonic string. 
These coincidences indicate that, despite lack of controlled
higher-order radiative corrections, the bosonic M(atrix) string theory
may offer a second-quantized description of interacting bosonic strings.

\vskip0.5cm
I am grateful to R. Dijkgraaf for collaboration on part of the results,
and to E. D'Hoker, L. Susskind, H. Tye and E. Witten for invaluable
discussions.

%%%%%%%%%%%%%%%%%%%%%%%%%%%%%%%%%%%%%%%%%%%%%%%%%%%%%%%%%%%%%%%%%%%

\end{document}